\documentstyle[aps,twocolumn]{revtex}
\draft
\begin{document}
\author{Yan V. Fyodorov$^{}$\cite{leave}  and Hans-J\"{u}rgen Sommers}
\address{ Fachbereich Physik,
Universit\"{a}t-Gesamthochschule Essen\\
 Essen 45117, Germany}
\title{Parametric correlations of scattering phase shifts and  
fluctuations of
 delay times in few-channel chaotic scattering}

\date{\today}
\maketitle

\begin{abstract}
By using the supersymmetry method we derive an explicit expression  
for the parametric correlation function of densities of eigenphases  
$\theta_a$ of the S-matrix in a chaotic quantum system
with broken time-reversal symmetry coupled to continua via M  
equivalent open channels;$\,\,a=1,..,M$ .We use it to find the  
distribution of derivatives of these eigenphases over the energy  
("phaseshift times") as well as over an arbitrary external  
parameter.We also find the
 parametric correlations of Wigner-Smith delay times.
\end{abstract}

\narrowtext
				
 The most fundamental object characterizing the process of quantum  
scattering is the unitary S-matrix relating the amplitudes of waves  
incoming onto the system and the amplitudes of scattered (outgoing)  
waves.

Recently there was a considerable growth of interest in the  
phenomenon of quantum chaotic scattering ( see reviews  
\cite{Smilansky},\cite{Gasp}), both  
theoretically\cite{BS,Lew,Mello,Stone},and  
experimentally\cite{Marcus,cav}.
Due to the chaotic nature of the underlying scattering dynamics the  
S-matrix characteristics behave in an irregular way when parameters  
of either incoming waves (e.g. the energy)
or characteristics of the scattering region (e.g. the form and  
strength of the scattering potential, the strength of the magnetic  
field through the ballistic microstructure, etc.) are slightly  
changed. Because of this fact it seems to be most adequate to  
describe such a behaviour in terms of some statistical measures:  
distributions and correlation functions.

At present, there are two complementary theoretical tools employed  
to study open quantum systems whose closed classical counterparts  
demonstrate chaotic behaviour. These are the semiclassical approach  
\cite{Smilansky,Gasp,BS} and the stochastic approach  
\cite{Lew,VWZ,Mello}. The latter one is based on the notion of the  
universality of the chaotic scattering phenomenon.

 It is well known
that the majority of "closed" chaotic quantum systems of quite  
different microscopic nature shows a great degree of universality on  
the level of its  statistical characteristics of spectra and  
eigenfunctions stemming from the very basic fact of chaotic internal  
motion. Because of this universality one achieves the correct  
description of the properties of such systems by exploiting the  
similarity with ensembles of large random matrices (RM). The latter  
principle is a commonly accepted one in the domain of Quantum Chaos  
\cite{Bohigas}, see also \cite{AlSi}, and recently some sound  
arguments were given in favour of its general validity \cite{AAA}.

Provided the statistics of the system  Hamiltonian is specified,  
one can work out the S-matrix by standard methods in the theory of  
quantum scattering
\cite{MW} and study its statistical characteristics. This way was  
pioneered by Verbaarschot et al. \cite{VWZ}
who calculated the correlation function of S-matrix elements. Other  
characteristics of the S-matrix and related
quantities can be efficiently studied
in the framework of this approach as well \cite{Macedo,new,FS,Lehmann}.

 One can also try to make use of the expected universality directly  
on the level of S-matrix without any reference to the system  
Hamiltonian.
Such a method was developed in a great detail in a series of
papers by Mello and collaborators \cite{Mello}. The probability  
density for the whole S-matrix can be obtained if one makes the  
assumption of minimal information
 content of such a distribution respecting the requirements of  
S-matrix unitarity
and symmetry constraints. Provided all the relevant information  
about the
system is encoded in a value of the average S-matrix $\langle S \rangle$
the probability $P(S)$ was shown to be given by a so-called Poisson's
kernel \cite{Mello}. More recently, Brouwer \cite{Brouwer} has shown
that the general Poisson kernel can be derived directly from the  
Hamiltonian approach
if one takes the Hamiltonian matrix from a quite specific  
Lorentzian RM ensemble. In view of the asymptotic equivalence of the  
Lorentzian and Gaussian matrices \cite{Brouwer} it is natural to  
expect that the Poisson's kernel is of the same universality as the  
spectral statistics for closed systems discussed above.

If one wishes to study the dependence of S-matrix on external  
parameters  without explicitly considering the system Hamiltonian,  
one should make some additional statistical assumptions beyond the  
minimum information approach. One possible way is to simulate such a  
dependence by
a kind of "Brownian motion" in the corresponding S-matrix space.
Quite a detailed presentation of such a method with some attempts  
to find a microscopic justification of the whole procedure can be  
found in \cite{FPich}.
It turns out, however, that the Brownian motion picture is in  
disagreement
with the results obtained starting from the Hamiltonian formalism.  
Therefore, the latter approach seems to be the only consistent one  
when we are
interested in the parametric variations.

Apart from the S-matrix elements, the scattering phase shifts  
$\theta_a$ (defined
via the S-matrix eigenvalues $\exp{i\theta_a}$) are used  
intensively to characterize the chaotic scattering, see e.g  
\cite{BS}. Quite recently their statistical characteristics were  
studied numerically in some detail  \cite{JPich,SZZ}. The  
derivatives of phase shifts over the energy  
$\tau_a=\partial\theta_a/\partial E$ ( which we suggest to call the  
"phaseshift times" ) are particularly interesting and related to the  
mean time spent by a particle in the interaction domain
(Wigner-Smith delay time,see \cite{Lehmann} and references therein)  
as $\tau_w=\frac{1}{M}\sum \tau_a$.
The fluctuations of the individual phaseshift times $\tau_a$ can be  
therefore used to characterize the possible fluctuations of delay  
times in the scattering process. Let us mention that various aspects  
of the general problem of delay time characteristics is under quite  
intensive study for a while \cite{new,delay,Lehmann}.
 More general parametric derivatives of the scattering phase shifts  
can also be related to some observable quantities. As a particular  
example we mention the relation between the persistent currents and
the derivative of the total phase shift over the magnetic flux \cite{Ak}.

In order to be able to study statistics of phase shifts $\theta_a$  
within the Hamiltonian approach, it is convenient to use the  
following representation for the scattering matrix (see  
e.g.\cite{Brouwer}):
\begin{equation}\label{init}
\hat{S}=[\hat{I}-i\pi W^{\dagger}\hat{R}(E)W]\times [\hat{I}+i\pi  
W^{\dagger}\hat{R}(E)W]^{-1}
\end{equation}
where we introduced the resolvent $\hat{R}(E)=(E-{\cal H})^{-1}$  
corresponding to the Hamiltonian ${\cal H}$ of the "closed" chaotic  
system and $W$ is the
matrix that couples the chaotic region with the incoming/outgoing waves.
In the stochastic approach the system Hamiltonian ${\cal H}$ is  
replaced by the $N\times N$ matrix $H_{ij}$ taken from the  
corresponding Gaussian RM ensemble.
In order to allow for the dependence of the S-matrix on the  
external parameter $X$ to be taken into account it is convenient to  
consider the set of RM $H_{ij}$ to be of the form \cite{AlSi}
: $\hat{H}(X)=\hat{H}_{0}+\frac{X}{\sqrt{N}}\hat{H}_1$.
To be specific, we consider in the present paper the simplest  
possible case when $\hat{H}_0$ runs over the Gaussian Unitary  
Ensemble (GUE) and $\hat{H}_1$
is an arbitrary but {\it fixed} matrix from the same ensemble.  
Physically this case is known to describe systems with completely  
broken time reversal symmetry (TRS). The scattering in systems with  
partly or completely broken TRS is under quite intensive  
theoretical\cite{Stone,FPich,JPich,theo} and
experimental \cite{TRS} study.

  The matrix $W$ is $N\times M$ matrix of amplitudes $W_{ia},\,\,  
a=1,2,...,M$, coupling the internal motion to $M$ open channels.  
Without much loss of generality these amplitudes can be chosen fixed  
in a way ensuring that the average
$S-$matrix is diagonal in the channel basis:  
$\overline{S_{ab}}=\delta_{ab}\overline{S_{aa}}$ \cite{VWZ}.   
Provided the energy $E$ is real, one finds the following expression  
\cite{VWZ}:
\begin{equation}\label{sav}
\overline{S_{aa}}=\frac{1-\gamma g(E)}{1+\gamma g(E)};\quad  
\gamma=\pi\sum_{i}W^{*}_{ia}W_{ia}
\end{equation}
where $g(E)=i E/2+(1-E^2/4)^{1/2}$ and we assumed that all $M$  
channels are statistically equivalent for the sake of simplicity.    
The strength of coupling to continua is convenient to be  
characterized via the "sticking probabilities"
(also called the "transmission coefficients" \cite{Lew,VWZ})
$T_{a}=1-|\overline{S_{aa}}|^2$ which are given for the present  
case by the following expression:
\begin{equation}\label{trans}
T_{a}^{-1}=\frac{1}{2}
\left[1+\frac{1}{2 \mbox{Re}\, g(E)}(\gamma+\gamma^{-1})\right]
\end{equation}
 The quantity  $T_{a}$ measures the part of the  flux in channel  
$a$ that spends substantial part of the time in the interaction  
region\cite{Lew,VWZ}.
We see that both limits
$\gamma\to 0$ and $\gamma\to \infty$ equally correspond to the weak  
effective coupling regime $T_{a}\ll 1$ whereas the
strongest coupling (at fixed energy $E$ ) corresponds to the value  
$\gamma=1$. The maximal possible coupling corresponding to the upper  
bound  $T_a=1$ is achieved in the present model for an energy  
interval in the vicinity of the center $E=0$.

The eq.(\ref{init}) shows that eigenphases $\theta_a$ considered  
modulo $2\pi$
\cite{note} are determined in a unique way by the eigenvalues  
$z_a(E,X)$ of the matrix $\hat{A}_X(E)=\pi W^{\dagger}\hat{R}_X(E)W$  
in view of the relation: $\theta_a=
-2\arctan{z_a}$. To this end let us introduce the density $\rho_{E,X}(z)=
\frac{1}{M}\sum_{a=1}^{M} \delta(z-z_a(E,X))$ and consider the  
correlation function:
\begin{equation}\label{defcor}\begin{array}{c}
{\cal K}_{E,\Omega,X}(z_1,z_2)=\\ \langle  
\rho_{E,0}(z_1)\rho_{E+\Omega,X}(z_2)
\rangle- \langle \rho_{E,0}(z_1)\rangle\langle\rho_{E+\Omega,X}(z_2)
\rangle\end{array}
\end{equation}
where the angular brackets stand for the averaging over the RM ensemble.
This correlation function can be easily found provided the  
following one is known for $\epsilon\to 0^{+}$:
\begin{equation}\label{gen1}
f(z_1,z_2)=\left\langle \mbox{Tr}\frac{1}{z_1-i\epsilon-\hat{A}_{0}(E)}
\mbox{Tr}\frac{1}{z_2+i\epsilon-\hat{A}_{X}(E+\Omega)}\right\rangle
\end{equation}

A somewhat unpleasant feature of eq.(\ref{gen1}) is that the random  
matrix
$\hat{H}$ enters it only via the matrix $\hat{A}$. However, due to the
identity
\begin{eqnarray}\label{iden}
\mbox{det}(z-\hat{A}_X(E))=\\ \nonumber
z^M \mbox{det}[ \hat{R}_X(E)(E-\hat{H}(X)-\frac{\pi}{z} WW^{\dagger})]
\end{eqnarray}
 the right hand side of eq.(\ref{gen1}) can be written in the  
following form:
$$ \begin{array}{c}f(z_1,z_2)=\frac{\partial^2}{\partial J_1\partial J_2}
\left[\left(\frac{Z_J^{(1)}Z_J^{(2)}}{Z_{J=0}^{(1)}Z_{J=0}^{(2)}}\right)^M
{\cal F}(J_1,J_2)\right]\mid_{J_1=J_2=0} \\ {\cal F}(J_1,J_2)=
\left\langle\frac{\mbox{  
det}[E-H_{eff}(0;Z_J^{(1)})]\mbox{det}[E+\Omega-H_{eff}(X;Z_J^{(2)})]}
{\mbox{det}[E-H_{eff}(0;Z_{J=0}^{(1)})]\mbox{det}[E+\Omega-H_{eff}(X;Z_{J=0}^{(2)})]}\right\rangle\end{array}  
$$
where we introduced the notations:  
$Z_J^{(p)}=z_p-i(-1)^p\epsilon+J_p;\quad
p=1,2$ and  
$H_{eff}(X;Z_J^{(p)})=\hat{H}(X)+\frac{\pi}{Z_J^{(p)}}WW^{\dagger}$.
In the latter form the expression is very close in its structure to that
used to study parametric correlations in ensembles of large RM  
\cite{AlSi}
and can be evaluated by essentially the same method. The general  
strategy is as follows: 1) to represent determinants in the  
denominator of the preceeding expression by auxilliary Gaussian  
integrals over commuting variables and those in the numerator by  
similar integrals over anticommuting variables. 2) To perform the  
averaging over the GUE distribution of $H_0$. 3) After employing the  
Hubbard-Stratonovich transformation to exploit the limit $N\gg M$  
when integrating out some ("massive") degrees of freedom in the  
saddle-point
approximation. After this is done the integral over the remaining  
("massless") degrees of freedom can be represented in a form of the  
so-called
 zero-dimensional supermatrix nonlinear $\sigma-$model introduced
for the first time by Efetov \cite{susy}.

For the {\it connected} part of the correlation function  
eq.(\ref{defcor}), we find after the set of standard manipulations  
\cite{VWZ,AlSi} the following expression:
\begin{eqnarray}\label{main}
f(z_1,z_2)=\int_{-1}^{1}d\lambda\int_{1}^{\infty}d\lambda_1  
\frac{{\cal F}_M(\lambda,\lambda_1)}{(\lambda_1-\lambda)^2}\times
\\ \nonumber \exp\{i\omega(\lambda_1-\lambda)-\frac{x^2}{2}
(\lambda_1^2-\lambda^2)\}.  \end{eqnarray}
\begin{eqnarray}\nonumber
{\cal F}_M(\lambda_1,\lambda)=\frac{\partial^2}{\partial z_1\partial z_2}
\left[\frac{z_1z_2+\gamma^2+i\gamma(z_1-z_2)\lambda}
{z_1z_2+\gamma^2+i\gamma(z_1-z_2)\lambda_1}\right]^M
\end{eqnarray}
where we put for simplicity the energy parameter $E=0$ and made the  
natural rescaling \cite{AlSi}: $\omega=\pi\Omega/\Delta;\quad  
x^2=(\pi X/N\Delta)^2\frac{1}{N}\mbox{Tr}\hat{H}_1^2$, with  
$\Delta\propto 1/N$ being the mean
level spacing of the GUE at $E=0$.

The eq.(\ref{main}) describes the parametric correlations of
eigenvalues of the matrix $\hat{A}_X(E)$ in closed form and  
provides the basis for extracting the statistical properties of  
scattering phase shifts.
First of all, it is evident that the integrations in eq.(\ref{main})
can be performed explicitly when $x=\omega=0$ thus giving the expression
for the correlation function of phase shifts at fixed value of the  
energy $E$ and the parameter$ X$. In order to represent it in a  
simple form for any value of
the coupling constant $\gamma$ it is convenient to introduce new  
variables
$\phi$ related to the scattering phase shifts $\theta$ as $\phi=
\arctan\{\gamma^{-1}\tan{(\theta/2)}\}$. A direct calculation shows  
that the correlation function
of the densities of the angles $\phi$ is equal to
\begin{equation}\label{corphi}
{\cal  
K}(\phi_1,\phi_2)\mid_{\phi_1\ne\phi_2}=-\left(\frac{\sin{M(\phi_1-\phi_2)}}
{\pi M\sin{(\phi_1-\phi_2)}}\right)^2
\end{equation}
for any number of open channels. Remembering the relation between  
$\gamma$ and $\langle S\rangle$, see eq.(\ref{sav}),
 one can satisfy oneself that eq.(\ref{corphi}) exactly coincides  
with the pair correlation function following from the Poisson's  
kernel distribution as defined in \cite{Mello,Brouwer}.

Let us turn our attention to the distributions of the individual  
phaseshift times $\tau_a=\partial \theta_a/\partial E$ and general  
parametric derivatives of phase shifts.  In view of the relation  
$\theta_a=-2\arctan{z_a}$ these distributions can be found easily  
provided the joint probability densities ${\cal P}_E(z,t)$ and  
${\cal P}_X(z,y)$ of $z_a$ and one of its parametric derivatives --  
$t_a=\partial z_a/\partial E$ or $y_a=\partial z_a/\partial X$ -- is  
known.
The functions ${\cal P}_E$ and ${\cal P}_X$ can be determined from  
the correlation function eq.(\ref{defcor}) as follows (see e.g.  
\cite{KZ}):
\begin{eqnarray} \label{deftau}\begin{array}{c}
{\cal P}_E(z,t)=\frac{1}{M}\left\langle  
\sum_{a=1}^{M}\delta(z-z_a)\delta(t-\partial z_a/\partial E)\right
\rangle\\=M\lim_{\Omega\to 0} \Omega {\cal  
K}_{E,\Omega,X=0}(z_1=z,z_2=z+t\Omega)\end{array} \\  
\begin{array}{c}
{\cal P}_X(z,y)=\frac{1}{M}\left\langle  
\sum_{a=1}^{M}\delta(z-z_a)\delta(y-\partial z_a/\partial X)\right
\rangle\\=M\lim_{X\to 0} X {\cal K}_{E,\Omega=0,X}(z_1=z,z_2=z+yX)  
\end{array}
\end{eqnarray}

Using the procedure described above we find the following  
distribution of
{\it inverse} (scaled) phaseshift times $u=2\pi/(\tau\Delta)$:
\begin{eqnarray}\label{disu}
{\cal  
P}_u(u)\equiv\frac{1}{M}\left\langle\sum_{a=1}^{M}\delta[u-2\pi/(\tau_a\Delta)]
\right\rangle\\
\nonumber
=\frac{(-1)^M}{M!}u^M\frac{d^M}{d  
u^{M}}\left(e^{-gu}I_{0}(u\sqrt{g^2-1})\right)
\end{eqnarray}
where $I_0(z)$ stands for the modified Bessel function,  
$g=2T^{-1}-1$ and $T$ is the "sticking probability" defined earlier.
Similarly, the scaled parametric derivatives  
$v_a=\partial{\theta_a}/\partial{x}$ are found to be distributed  
according to the probability density
\begin{equation}\label{disv}
{\cal P}_v(v)=\int_0^{\infty}\frac{du \,\,u}
{(2\pi)^{1/2}}\exp{\left[-\frac{u^2v^2}{2}\right]}{\cal P}_u(u)
 \end{equation}
where ${\cal P}_u(u)$ is defined in the preceeding equation. Both  
distributions depend on the coupling constant
$\gamma$ only via the sticking probability $T$ rather than via the  
average
S-matrix. They assume the simplest form for the "critical" coupling  
$T=1$ corresponding to the most strong overlap of individual  
resonances allowed for the few-channel scattering \cite{new,FS}.  
Under this condition one finds the following distribution of scaled  
phaseshift times $\tau_s=1/u$:
 ${\cal P}_\tau(\tau_s)=a=\frac{1}{M!}\tau_s^{-M-2}e^{-1/\tau_s}$.

 In the recent paper \cite{SZZ} this formula was found to be
in a good agreement with results of direct numerical simulations.
The authors also found a way to arrive
at such an expression exploiting some plausible assumptions about the
statistics of S-matrix eigenphases and their derivatives.

Having at our disposal the exact distribution eq.(\ref{disu})
it is instructive to calculate the mean value and the variance of the 
phaseshift times. One finds:
\begin{equation}\label{mom}
\langle \tau\rangle=\frac{2\pi}{M\Delta};\quad \frac{\langle  
\tau^2\rangle-\langle \tau\rangle^2}{\langle \tau\rangle^2}=\frac{2M  
(T^{-1}-1)+1}{M-1}
\end{equation}

The first of these relations is quite well known  
\cite{delay,Lehmann}. It shows that the mean delay time $\langle  
\tau\rangle$ is determined
by the mean level spacing $\Delta$ of the closed system and the  
number $M$ of open channels. On the other hand the magnitude of  
delay time fluctuations
measured by the relative variance of the phaseshift time  
distribution, see eq.(\ref{mom}), is determined both by $M$ and $T$.  
Generically, the fluctuations are the weaker the larger is the  
number of open channels $M$ and the stronger is the
coupling to continua: $1-T\ll T$. Let us also mention as an  
interesting feature
the divergency of the time-delay variance
at $M=1$, which is a consequence of the powerlaw
tail $\tau_s^{-M-2}$
typical for the distribution ${\cal P}_{\tau}(\tau_s)$.

Eq.(\ref{main}) can be used to study the parametric correlation
functions of the Wigner-Smith delay times $\tau_w(E,X)$. However,  
this calculation
turns out to be quite lengthy and will be presented in a more  
extended publication. On the other hand, remembering the definition
$\tau_W=-\frac{i}{M}
\frac{\partial}{\partial E}\ln{\mbox{det} S(E)}$ one finds the relation 
$
\tau_w=-\frac{2}{M}\mbox{Im Tr}(E-{\cal H}+i\pi WW^{\dagger})^{-1}
$
which follows directly from eq.(\ref{init}) upon using the identity
eq.(\ref{iden}).
Now it is evident that the calculation of the correlation of  
fluctuations of Wigner-Smith delay times $\delta\tau_{W}(E,X)=
\tau_W-\langle\tau_W\rangle$ amounts to evaluating the average  
product of the resolvents
of the non-Hermitian effective Hamiltonians ${\cal H}\pm i\pi WW^{+}$. 
For the case of chaotic systems with preserved TRS such a  
calculation was done  earlier in \cite{Lehmann}. For the present  
case we find:
\begin{eqnarray}\label{Wig} \nonumber
\langle\delta\tau_{W}(E,0)\delta\tau_{W}(E+\Omega,X)\rangle= \\
2\left(\frac{\pi}{M\Delta}\right)^2\times  
\int_{-1}^{1}d\lambda\int_{1}^{\infty}d\lambda_1  
\cos{\omega(\lambda_1-\lambda)} \\ \nonumber
\exp\{-\frac{x^2}{2}
(\lambda_1^2-\lambda^2)\}
\times \left[\frac{1+\lambda\,g^{-1}}{1+\lambda_1\,g^{-1}}\right]^M
\end{eqnarray}
 The variance of the $\tau_w$
is given by:
\begin{equation}
 \frac{\langle \tau_W^2\rangle-\langle \tau_W\rangle^2}{\langle  
\tau_W\rangle^2}=\frac{2}{T^2(M^2-1)}\left[1- (1-T)^{M+1}\right]
\end{equation}
showing the same qualitative features ( divergencies at
$M=1$ or $T\to 0$) as those following from the eq.(\ref{mom}).

In conclusion, we studied analytically the parametric correlations  
of scattering phase shifts and the distribution of phase shift  
derivatives for a
chaotic quantum  system with broken TRS. The method of calculation
 automatically
provides the universality of the obtained  results for a very broad  
class of
systems which can be described by the
mapping on the same supermatrix nonlinear $\sigma-$ model  
\cite{AlSi}. Another important feature of the method is that it  
provides the regular
basis for studying a general case of disordered systems where the  
effects of localization modify essentially the phaseshift statistics  
\cite{JPich}.
This issue as well as the extension of our results to the case of  
systems with preserved TRS are left for future research.

Authors are very obliged to P.Seba,J.Zakrzewski and K.Zyczkowski  
for providing them with the text of the paper\cite{SZZ} prior to  
publication. Y.V.F. appreciates  stimulating and clarifying  
discussions on chaotic scattering with P.Seba and is grateful to  
N.Lehmann for useful discussions.  The financial support by SFB 237  
"Unordnung und grosse Fluktuationen"
is acknowledged with thanks.

\end{document}